\documentclass[%
 reprint,
 amsmath,amssymb,
 aps,
]{revtex4-1}

\usepackage{graphicx}
\usepackage{dcolumn}
\usepackage{bm}
\usepackage{braket}

\newcommand{\Tc}{$T_\mathrm{c}$ }
\newcommand{\Jc}{$J_\mathrm{c}$ }

\newcommand{\sample}{BaFe$_{2}$(As$_{0.67}$P$_{0.33}$)$_2$ }
\newcommand{\Jcm}{J_\mathrm{c} }
\graphicspath{{./fig-intro/}}

\newcommand{\BaCox}{Ba(Fe$_{1-x}$Co$_x$)$_2$As$_2$ }

\newcommand{\Tcm}{T_\mathrm{c}}

\begin{document}

\preprint{APS/123-QED}

\title{Effects of 6 MeV proton irradiation on the vortex ensemble in BaFe$_{2}$(As$_{0.67}$P$_{0.33}$)$_2$ \\ revealed through magnetization measurements and real-space vortex imaging}
\author{Akiyoshi Park}
\author{Ivan Veshchunov}
\author{Akinori Mine}
\author{Sunseng Pyon}
\author{Tsuyoshi Tamegai}
\affiliation{Department of Applied Physics, The University of Tokyo, Hongo, Bunkyo-ku, Tokyo 113-8656, Japan\\}

\author{Hisashi Kitamura}
\affiliation{Center for Advanced Radiation Emergency Medicine, National Institutes for Quantum and Radiological Science and Technology, 4-9-1, Anagawa, Inage-ku Chiba, 263-8555, Japan}


\date{\today}

\begin{abstract}
The change in vortex ensemble in BaFe$_{2}$(As$_{0.67}$P$_{0.33}$)$_2$, an isovalently doped iron-based superconductor (IBS), is studied through global magnetization measurements and single vortex imaging before and after 6 MeV proton irradiation. The field dependence of the critical current density ($J_\mathrm{c}$) is analyzed through the strong pinning model, with which the pristine sample is consistent.  After the irradiation, the $J_\mathrm{c}$ aberrates from the strong pinning field dependence of $B^{-5/8}$, and evolves to a weaker $B^{-1/3}$ dependence with an anomalous two-step behavior creating a cusp like feature.  The cusp coincides with the field of the local minima in the normalized relaxation rate ($S$), manifested by increased pinning due to increased intervortex interactions followed by fast vortex dynamics caused by flux activation at higher fields. Furthermore, single vortex imaging reveals that while long-range triangular correlation of the Abrikosov vortex lattice is observed in pristine samples, irradiated samples exhibit a highly disordered glassy vortex state which is more densely packed with increased pinning force. These artificial defects incorporated via proton irradiation have the same pinning effect as Co doping which not only shifts the pinning force to a higher degree but also broadens its distribution, in contrast to P doping which only broadens the pinning distribution without inducing a shift. All in all, through this investigation, we provide a systematic understanding of the pinning behavior in BaFe$_{2}$(As$_{0.67}$P$_{0.33}$)$_2$ through carefully controlling the defects in the system.

\begin{description}
\item[PACS numbers]
74.70.Xa, 74.62.En, 74.25.fc 
\end{description}
\end{abstract}

\maketitle


\section{\label{sec:level1}Introduction}

Exhibiting robust superconductivity against disorder, iron-based superconductors (IBSs) demonstrate significant enhancement in the stabilization of vortices through incorporation of defects \cite{PhysRevB.80.012510, PhysRevB.81.094509, PhysRevB.81.094509, doi:10.1063/1.4731204, 0953-2048-25-8-084008}.  Yet, the understanding of the underlying pinning phenomena falls short due to the complexity of the dynamic process involving the combined contribution of pinning energy, line tension, and vortex-vortex interactions \cite{PhysRevLett.92.067009}.  To deconvolute the issue, it is thus important to control the effects of each of these factors through carefully managing the defects and the magnetic field in the system.  In particular, the difficulty lies in the ability to control the amount of defects in IBSs, as the pinning landscape in IBSs is manifested by various forms of defects inherent to each individual crystal. Ba$_{1-x}$K$_{x}$Fe$_{2}$As$_{2}$ single crystals in the underdoped regime have exhibited pinning dominated by twin boundaries \cite{PhysRevB.85.014524}. Moreover, Ba(Fe$_{1-x}$Co$_{x}$)$_{2}$As$_{2}$ single crystals have shown that pinning is dictated by nanometer scale spatial fluctuations in superfluid density caused by inhomogeneous doping \cite{PhysRevB.84.094517}.
Thus, it is our intention to systematically control the pinning in IBSs through synthesizing high-quality crystals with negligible intrinsic pinning, and incorporating defects through controlled irradiation of protons, then experimentally clarifying the pinning mechanism involved. 

Amongst the various types of particle irradiation, proton (H$^+$) irradiation has been reported to effectively enhance the critical current density (\Jc) through introducing point-defects as observed in Ba$_{1-x}$K$_{x}$Fe$_{2}$As$_{2}$ \cite{0953-2048-28-8-085003} and Ba(Fe$_{1-x}$Co$_{x}$)$_{2}$As$_{2}$ \cite{PhysRevB.86.094527}.  To observe solely the effect of irradiation, optimally doped isovalent BaFe$_{2}$(As$_{0.67}$P$_{0.33}$)$_2$, which resides in the clean-limit with long electron mean-free path as examined through de Haas-van Alphen effect, appeals as an appropriate subject \cite{PhysRevLett.104.057008}.  Small angle neutron scattering and Bitter decoration on \sample have revealed the existence of triangular lattice in this system, further asserting its highly ordered vortex ensemble \cite{PhysRevB.90.125116, PhysRevB.87.094506}.

In this paper, we explore the change in pinning mechanism with the introduction of disorder through 6 MeV proton irradiation in \sample superconductor by studying the change in the critical current density, investigating the dynamics of vortices, and through real space vortex imaging.  We stress that the high quality of the sample with small inhomogeneities, and small intrinsic defects, allows us to observe the pure effects of vortex pinning by defects dominated by particle irradiation.
\vspace{-10 pt}
\section{Experimental details}
\sample single crystals were grown through a Ba$_2$As$_3$/Ba$_2$P$_3$ flux method as described in Ref. \cite{doi:10.1143/JPSJ.81.104710}.  Pre-synthesized Ba$_2$As$_3$, Ba$_2$P$_3$, FeAs and FeP precursors were placed in an alumina crucible, and vacuum sealed in a quartz tube.  All procedures were carried out in a glove box with N$_2$ atmosphere in order to prevent oxidation of chemically unstable Ba$_2$As$_3$ and Ba$_2$P$_3$.  The assembly was heated to 1150 $^\circ$C and cooled to 900 $^\circ$C over a period of 250 h in an electric furnace, yielding crystals of several millimeters.  The employed method allows for synthesis of higher quality crystals over the method of crystal growth in the FeAs flux, since the Ba$_2$As$_3$/Ba$_2$P$_3$ precursors provided a phosphorous rich flux, thereby resulting in crystals with high homogeneity.
 
\begin{figure}[t]
\includegraphics[width=8 cm]{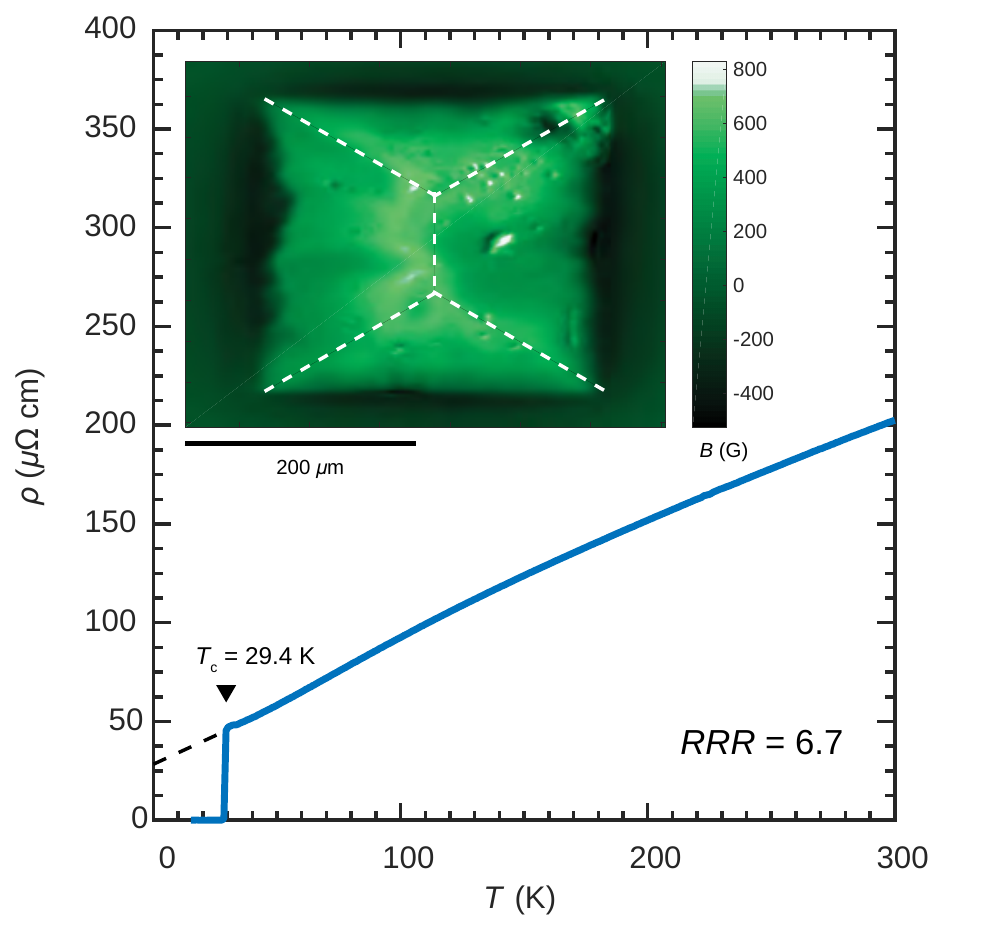}
\caption{The temperature dependence of resistivity in the pristine sample.  The residual resistivity ratio is 6.7. Inset: Magneto-optical imaging of \sample single crystal at the remanent magnetization state after sweeping to a field up to 1.6 kOe at $T = 20$ K. The white dashed lines outline the double Y feature of the discontinuity line. }
\label{RRR}
\end{figure}

\begin{figure}[t]
\includegraphics[width=8 cm]{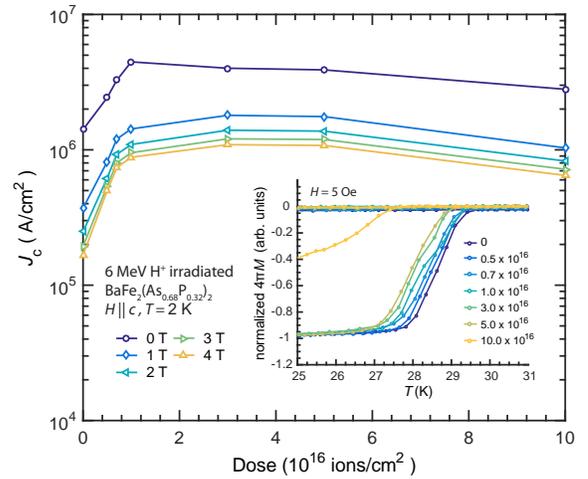}
\caption{\label{dose} Proton (H$^+$) dose dependence of critical current density ($\Jcm$) at 2 K. The inset indicates the change in the diamagnetic transition for each H$^+$ dose in which the diamagnetic signal was normalized by the maximum diamagnetic response in the Meissner state to remove the effects of different sample size and shape.  }
\label{dosedependence}

\end{figure}

\begin{figure}[b]
\includegraphics[width=8.5 cm]{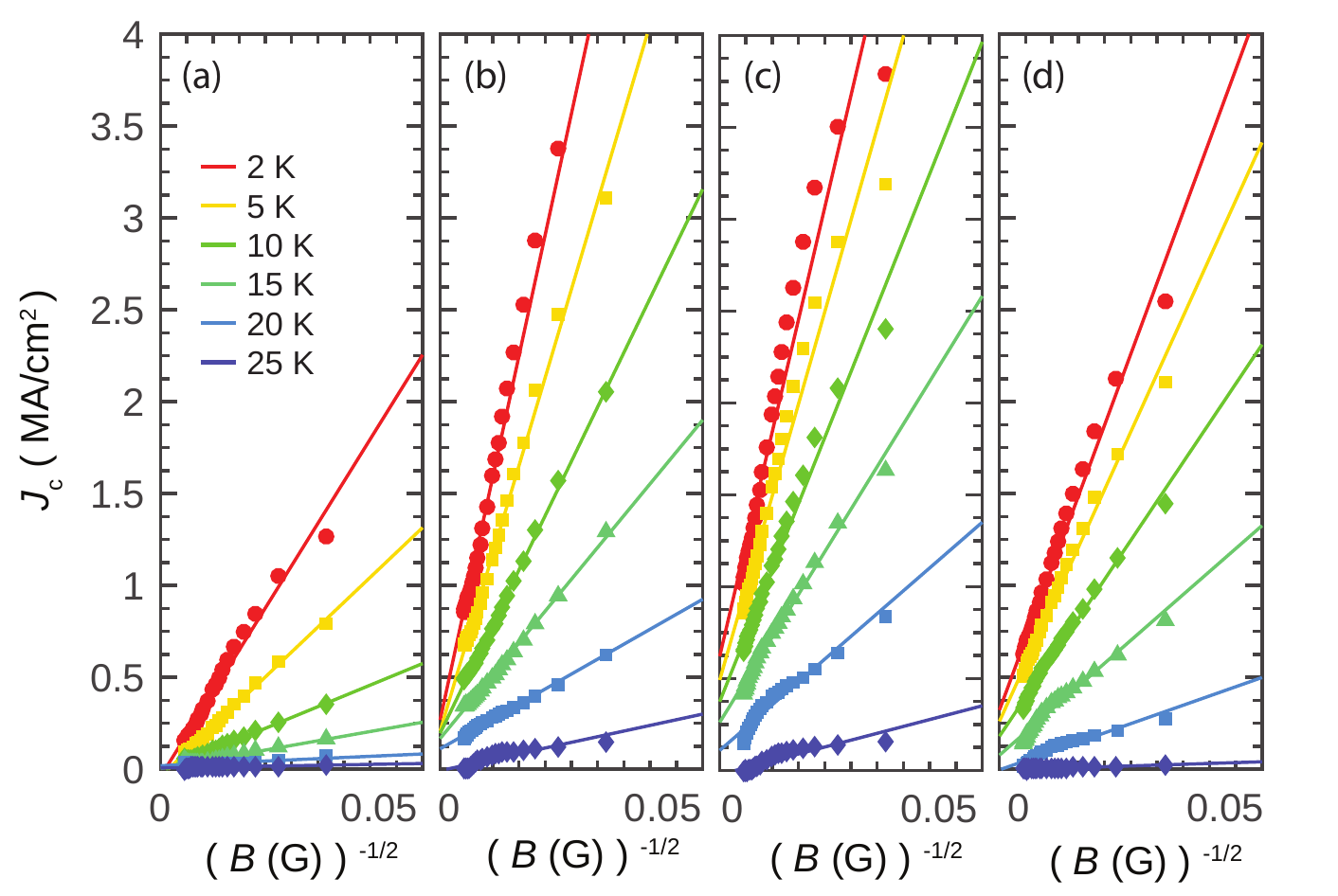}
\caption{The $\Jcm$ versus $1/\sqrt{B}$ in which the slope of the linear regression curve is used to calculate the elementary pinning force $f_{p,s}$, for (a) pristine, and H$^+$ irradiated \sample with dose of (b) $1 \times 10^{16}$ ions/cm$^2$}, (c) $5 \times 10^{16}$ ions/cm$^2$ and (d) $10 \times 10^{16}$ ions/cm$^2$.  $\hspace{40 pt}$
\label{fps}
\end{figure}

\begin{figure*}
\includegraphics[width=0.85\textwidth]{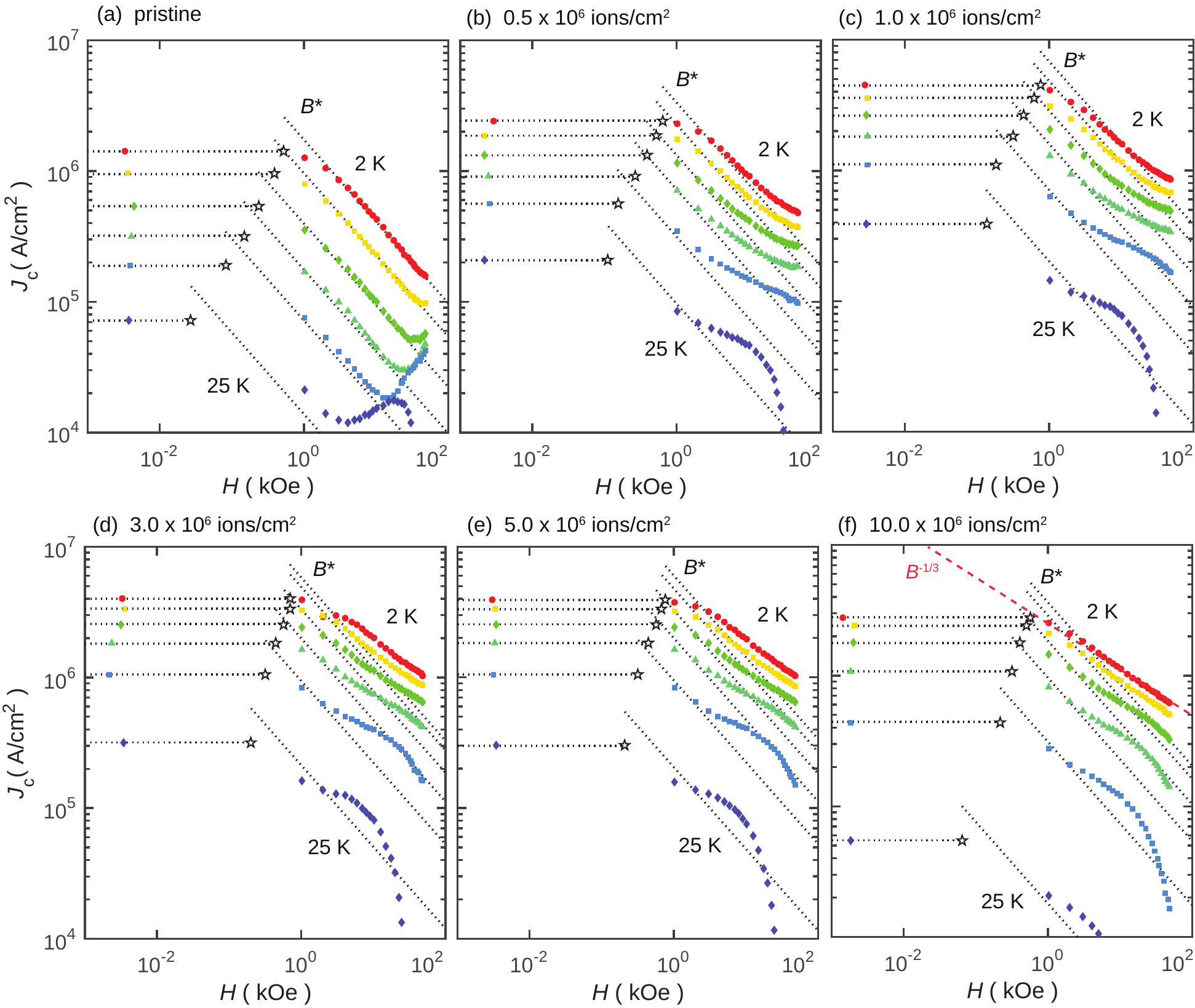}
\caption{Field dependence of \Jc for \sample with varying H$^+$ irradiation dose at different temperatures. The pentagram indicate the cross-over field $B^*$ and the dashed black lines indicate the \Jc proportional to $B^{-5/8}$ predicted at higher field regimes, from the strong pinning picture. The dashed red line in (f) indicates a $B^{-1/3}$ slope.}
\label{Jc}
\end{figure*}

The chemical stochiometry was determined through both energy dispersive x-ray analysis (EDX) and the angle of the ($00l$) peaks in the x-ray diffraction (XRD) pattern.  The superconducting transition temperature, $\Tcm$, was determined by the temperature of the resistivity drop, indicating  that \Tc$=$ 29.4 K, with a sharp transition width of $\Delta$\Tc$\approx 1$ K and small residual resistivity extrapolated to absolute zero, of approximately 20 $\mu\Omega$cm (Fig. \ref{RRR}), which is comparable to the residual resistivity of 20 $\mu\Omega$cm and 10 $\mu\Omega$cm reported in Refs. \cite{PhysRevB.84.140504, PhysRevB.98.054512} and Ref. \cite{10.1038/ncomms6657}.
 Furthermore, a high homogeneity of superconductivity on a $\mu$m-scale was evident from magneto-optical (MO) image (inset of Fig. \ref{RRR}) of the remanent state after applying a field of 1.6 kOe and returning back to zero-field.  The clear discontinuity line representing a double-Y structure is indicative of spatially homogeneous superconductivity. However, it should be noted that current discontinuity lines are distorted from the ideal double-Y structure due to the presence of macroscopic defects.

 Transmission electron microscopy has revealed that proton irradiation embodies spherical defects of nm-scale in \sample \cite{PhysRevB.98.054512}.  To allow for homogeneous incorporation of defects through proton irradiation, crystals were cleaved into a thickness of $<30$ $\mu$m. The crystals were mounted on a Al plate with Ag paste and irradiated along the $c$ axis at the National Institute of Radiological Sciences-Heavy Ion Medical Accelerator in Chiba (NIRS-HIMAC), located in Japan.  Samples were irradiated with 6 MeV protons at room temperature with different irradiation doses: 0.5, 0.7, 1.0, 3.0, 5.0, 10.0 $\times$ 10$^{16}$ ions/cm$^2$. In order to ensure that the crystals are identical in character, the samples subject to measurements were collected from the same batch.  Furthermore, although irradiation at room temperature may induce thermal annealing which reduces the actual number of defects from the nominal number of defects, we observed prominent changes in $\Jcm$,  and reduction in $\Tcm$, indicating appropriate embodiment of defects in crystals, thereby serving our purpose.

After the irradiation, crystals were subject to magnetization measurements performed by a commercial superconducting quantum interference device  (SQUID) magnetometer with an applied field along the $c$ axis.  Furthermore, Bitter decoration was performed with a crystal irradiated with a dose of $3\times 10^{16}$ ions/cm$^2$ together with the pristine one.  The sample was cleaved to create a clean fresh surface which was decorated with iron particles at liquid helium temperature (4.2 K) after being field-cooled in $H = 40$ Oe from above $\Tcm$.  The sample surface was then imaged through a scanning electron microscope (SEM) to determine the position of the vortices.  Furthermore, Bitter decoration was employed as a filter for distinguishing underdoped samples with twin domains.  Since the optimal-doping level resides in close vicinity of the antiferromagnetic phase, many of the samples are likely to have twin domains, which can contribute to additional pinning.  Since the purpose of this study is to see the pure effects of irradiation, samples with total absence of twin domains were carefully selected as a subject for the investigation.
\vspace{- 10 pt}
\section{Results}

\subsection{Critical current density}
In order to understand the change in vortex pinning brought about by proton irradiation, we measure magnetic hysteresis (\textit{M-H}) loops at various temperatures with various irradiation doses.  From the \textit{M-H} loops, the $\Jcm$ is calculated using the extended Bean model :  $J_\mathrm{c}=20\Delta M/a(1-a/3b)$ where $a$ and $b$ are the length and width of the rectangular crystal in centimeters in which $a < b$ and $\Delta M$ in the unit of G, is the width of the magnetization curve in the negative field regime when sweeping down field and returning back to zero field \cite{PhysRevLett.8.250}.  Note that, strictly speaking, the $\Jcm$ measured through magnetization measurements can be less than the actual $\Jcm$ due to finite magnetic relaxation, and the value obtained by the Bean model would yield the sustainable current ($J_\mathrm{s}$).  However, given that most publications have referred to the value as $\Jcm$, we have employed the same notation for consistency.

Focusing on the $\Jcm$ of the pristine crystal, it becomes evident that unlike the previous report of Ref. \cite{PhysRevLett.105.267002}, the crystals employed for this measurement indicate a clear presence of a nonmonotonic field dependence, often referred to as the fish-tail effect. The initially reported absence of the peak phenomena in \sample observed in other isovalently-doped systems, was interpreted as an outcome of the lack of charged dopants, leading to an understanding that the charged dopants are responsible for weak-collective pinning \cite{PhysRevLett.105.267002,1742-6596-449-1-012023}. However, the presence of the fish-tail observed here clearly undermines such an argument. In addition, Ba(Fe$_{0.64}$Ru$_{0.36}$)$_2$As$_2$, another isovalently doped IBS, exhibits a clear fish-tail effect, further supporting the recent result of Ref. \cite{Ohtake}.

From the MO images of the samples in Ref.\cite{PhysRevB.87.094506}, clearly, there is a high degree of spatial inhomogeneity in superconductivity on a $\mu$m-scale, which we could rationalize as to masking the fish-tail effect that should be present in the clean disorder-less state. It is noteworthy that the spatial homogeneity in superconductivity observed through magneto-optical images in our sample indicates a much more homogeneous superconductivity than that of Ref. \cite{PhysRevB.87.094506} (Fig. \ref{RRR}). 

Extracting the \Jc at 2 K  as shown in Fig. \ref{dosedependence}, it becomes clear that for all field ranges, there is growth in \Jc with increasing irradiation dose up to $1\times 10^{16}$ ions/cm$^2$, where the $\Jcm$ saturates to a nearly constant value. Irradiating \sample with $1\times 10^{16}$ ions/cm$^2$ induces a 3.2-fold enhancement from $1.4\times 10^{6}$ A/cm$^2$ to $4.5 \times 10^{6}$ A/cm$^2$ at self-field. At higher irradiation doses, point-like defects created by proton irradiation induces substantial quasiparticle scattering of Cooper pairs, as suggested by the drop in \Tc [Fig. \ref{dosedependence} (inset)], flattening out the enhancement in $\Jcm$ \cite{0953-2048-25-8-084008}.

In the strong pinning scenario with a density $n_\mathrm{i}$ of strong pins of sizes larger than the in-plane coherence length,  ($\xi_{ab}$), the strong pinning contribution to the $\Jcm$ ($\Jcm^s$) in the low-field regime manifests a plateau region followed by a power-law decrease at higher field regimes, adhering to the relationship \cite{PhysRevB.66.024523, PhysRevB.43.8024, PhysRevB.81.174517},

\begin{align}
\Jcm^s (0) &= \frac{\pi^{1/2}n_i^{1/2}j_0}{\varepsilon_\lambda} \bigg( \frac{f_{p,s}\xi_{ab}}{\varepsilon_0} \bigg)^{3/2} \label{a} && (B<B^*) \\
\Jcm^s (B) &\approx \frac{2 n_ij_0}{\varepsilon_\lambda^{5/4}\xi_{ab}^{1/2}}\bigg( \frac{f_{p,s}\xi_{ab}}{\varepsilon_0}\bigg)^{9/4} \bigg(\frac{\Phi_0}{B}\bigg)^{5/8} && (B>B^*). \label{b}
\end{align}
Here, 
$\varepsilon_\lambda=\lambda_{ab}/\lambda_{c}$ is the penetration depth anisotropy,
$j_0 = 4\varepsilon_0/3\sqrt{3}{\Phi_{0}}\xi_{ab}$ is the depairing current density, $\varepsilon_0 = (\Phi_0/4\pi\lambda_{ab})^2$ is the vortex line energy, in which the $\lambda_{ab}$ is the $ab$-plane penetration depth,  $\lambda_{c}$ is the $c$-axis penetration depth and $\Phi_0$ is the flux quantum, all in CGS units.  For this particular case, values $\varepsilon_\lambda=0.15$ \cite{PROZOROV2009582}, $\xi_{ab}=1.6$ nm \cite{PhysRevLett.105.267002}, and $\lambda_{ab}= 200$ nm  
were employed \cite{Hashimoto1554}.  Further, $f_{p,s}$, the elementary pinning force of an individual pin is computed from the field-dependence of the $\Jcm$.

\begin{align}
f_{p,s} & = \frac{\Phi_0^{3/2}\varepsilon_\lambda\Jcm^2(0)}{\pi}\bigg(\frac{\partial \Jcm(B)}{\partial B^{-1/2}}\bigg)^{-1} \label{d},
\end{align}

At low fields, in the single-vortex limit, $ \Jcm^s$ obeys a field-independent behavior and is later followed by a $B^{-5/8}$ field-dependence at higher fields.  Single-vortex pinning is therefore, the main contributing pinning mechanism that takes place due to crystal imperfections of nanoscale.  Such a boundary between the two field regimes is denoted as the cross-over field $B^*$ which can be expressed by the equation below \cite{PhysRevB.66.024523}.

\begin{align}
B^* & \equiv \pi\Phi_0n_i\bigg(\frac{U_p}{\varepsilon_0}\bigg) \\
& \approx 0.74\frac{\Phi_0}{\varepsilon_\lambda^2}\bigg(\frac{n_i}{\xi_{ab}}\bigg)^{4/5}\bigg(\frac{f_{p,s}\xi_{ab}}{\varepsilon_0}\bigg)^{6/5} \label{c}
\end{align}

\begin{figure}[t]
\includegraphics[width=8 cm]{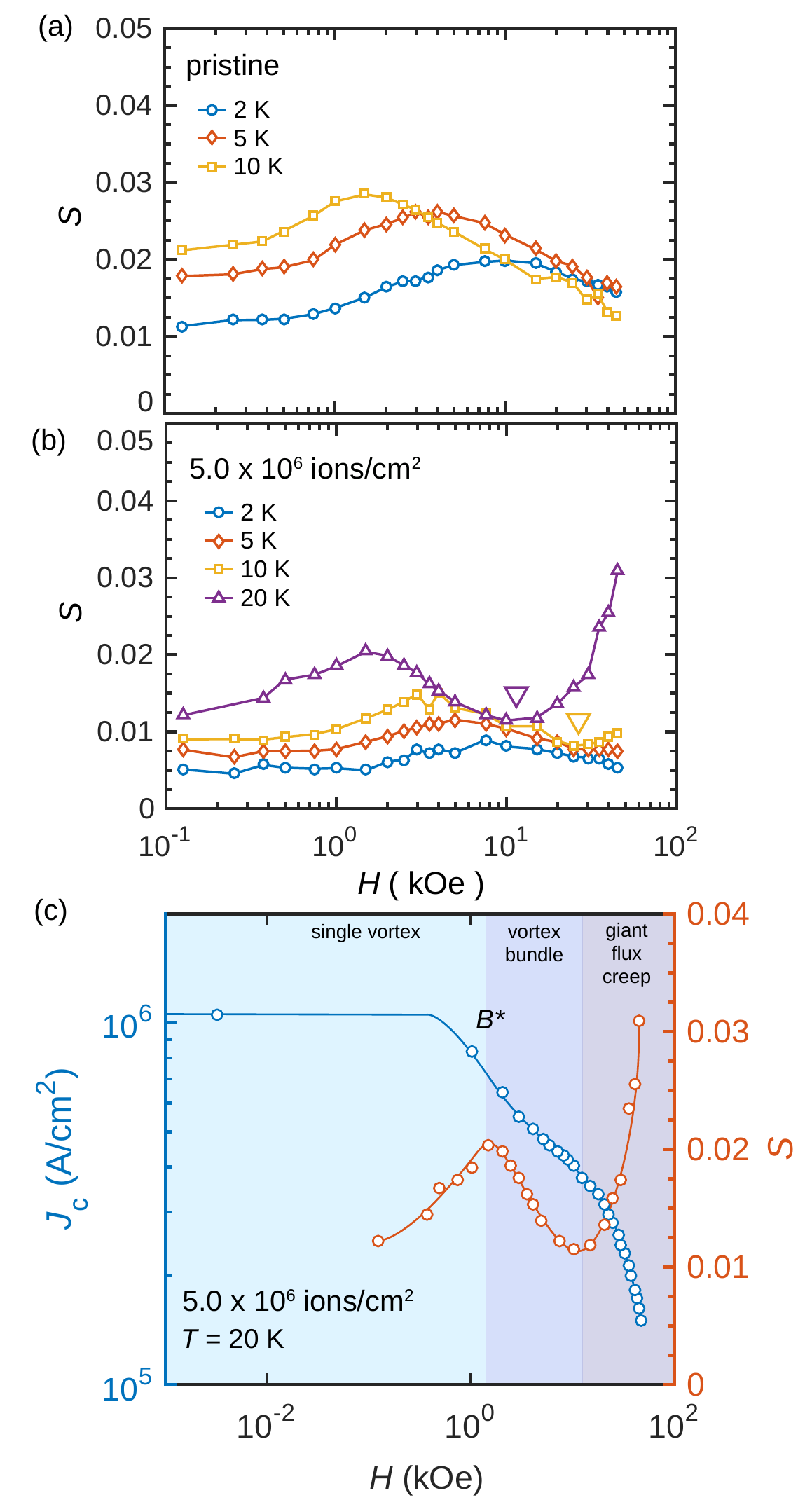}
\caption{The field dependence of the normalized relaxation rate in (a) pristine and (b) H$^+$ irradiated \sample with $5\times 10^{16}$ ions/cm$^2$. The $\bigtriangledown$ indicates the local minimum of the normalized relaxation rate, $S(H)$. (c) The $S(H)$ at 2 K and $\Jcm(H)$ superimposed on the same graph for \sample irradiated with $5\times 10^{16}$ ions/cm$^2$. }
\label{SH}
\end{figure}
\begin{figure*}[t]
\includegraphics[width=0.8\textwidth]{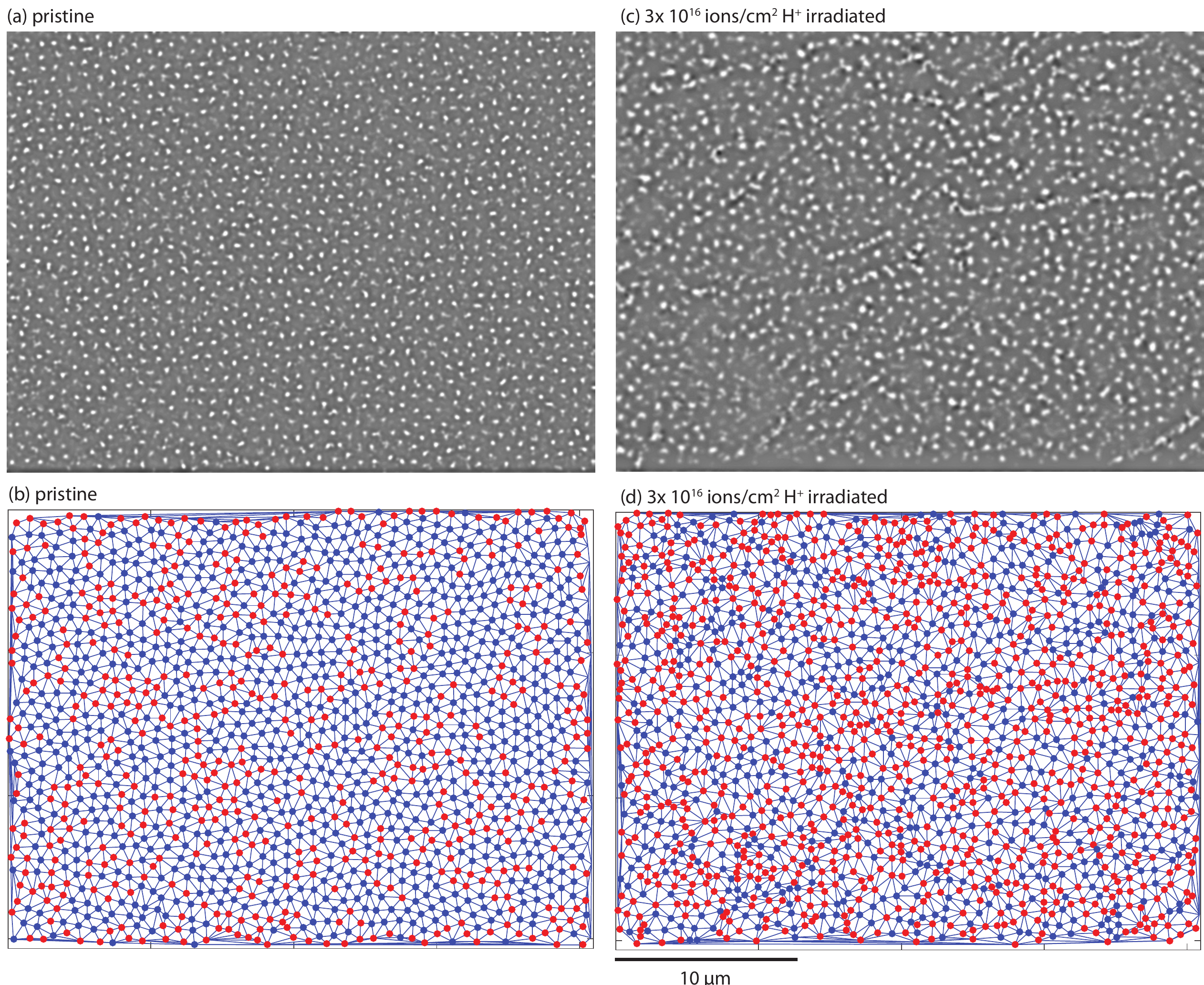}
\caption{\label{fig:epsart} Scanning electron microscopy images of decorated (a) pristine and (c) $3.0 \times 10^{16}$ ions/cm$^2$ H$^+$-irradiated BaFe$_{2}$(As$_{0.67}$P$_{0.33}$)$_2$ with applied field of 40 Oe. (b) and (d) mark the centers of the vortices and the Delaunay triangles.  Blue circles indicate vortices with a six-fold connectivity and red circles indicate vortices with a higher degree of connectivity.}
\label{bitterimage}
\end{figure*}

To assess how well the single-vortex pinning picture corresponds to experimental data, the experimentally obtained $\Jcm$ at the zero-field state is equated to the $\Jcm$ of the strong pinning field in the low-field regime such that, $\Jcm(0) = \Jcm^s(0)$, since the critical current density is manifested by strong pinning centers at the low-field regime, given that the vortex density would be too sparse for weak collective pinning to be contributing to the overall critical current density.  Upon this, $f_{p,s}$ is calculated by performing a linear regression of $J_\mathrm{c}(B)$ vs. $1/\sqrt{B}$ as shown in Fig. \ref{fps}, which indicates a linear relationship. The value of $f_{p,s}$ obtained in this way for every temperature is substituted to Eqs. \eqref{a}, \eqref{b}, \eqref{c} to compute $B^*$ and $\Jcm^s(B)$.

Figures \ref{Jc}(a)-\ref{Jc}(f) present the experimentally obtained \Jc at various temperatures along with $B^*$ and $\Jcm^s(B)$ calculated from the ansatz of the single-vortex pinning theory.  Clearly, the single-vortex pinning picture elegantly describes the \Jc of pristine \sample indicating a $B^{-5/8}$ dependence as predicted by the strong pinning theory.  Such power-law behavior has also been recognized in Ba$_{1-x}$K$_{x}$Fe$_{2}$As$_{2}$, Ba(Fe$_{1-x}$Co$_{x}$)$_{2}$As$_{2}$ and NdFeAsO$_{0.9}$F$_{0.1}$ \cite{0953-2048-28-8-085003, PhysRevB.86.094527, PhysRevB.81.174517}.
Furthermore, from the cross-over field $B^* [$Eq.\eqref{c}], and using the elementary pinning force $f_{p,s}$ [Eq.\eqref{d}], we solve for $n_i$, which allows us to estimate the approximate defect density $n_i$ ranging $0.5 \sim 3 \times 10^{14}$ cm$^{-3}$ in these crystals. This value is just an order of $10^6$ smaller that the concentration of atomic dopants, $8 \times 10^{20}$ cm$^{-3}$ \cite{PhysRevB.82.014513}, which in this case is the P atom. 

\begin{figure}[b]
\center
  \begin{center}
\includegraphics[width=7.5 cm]{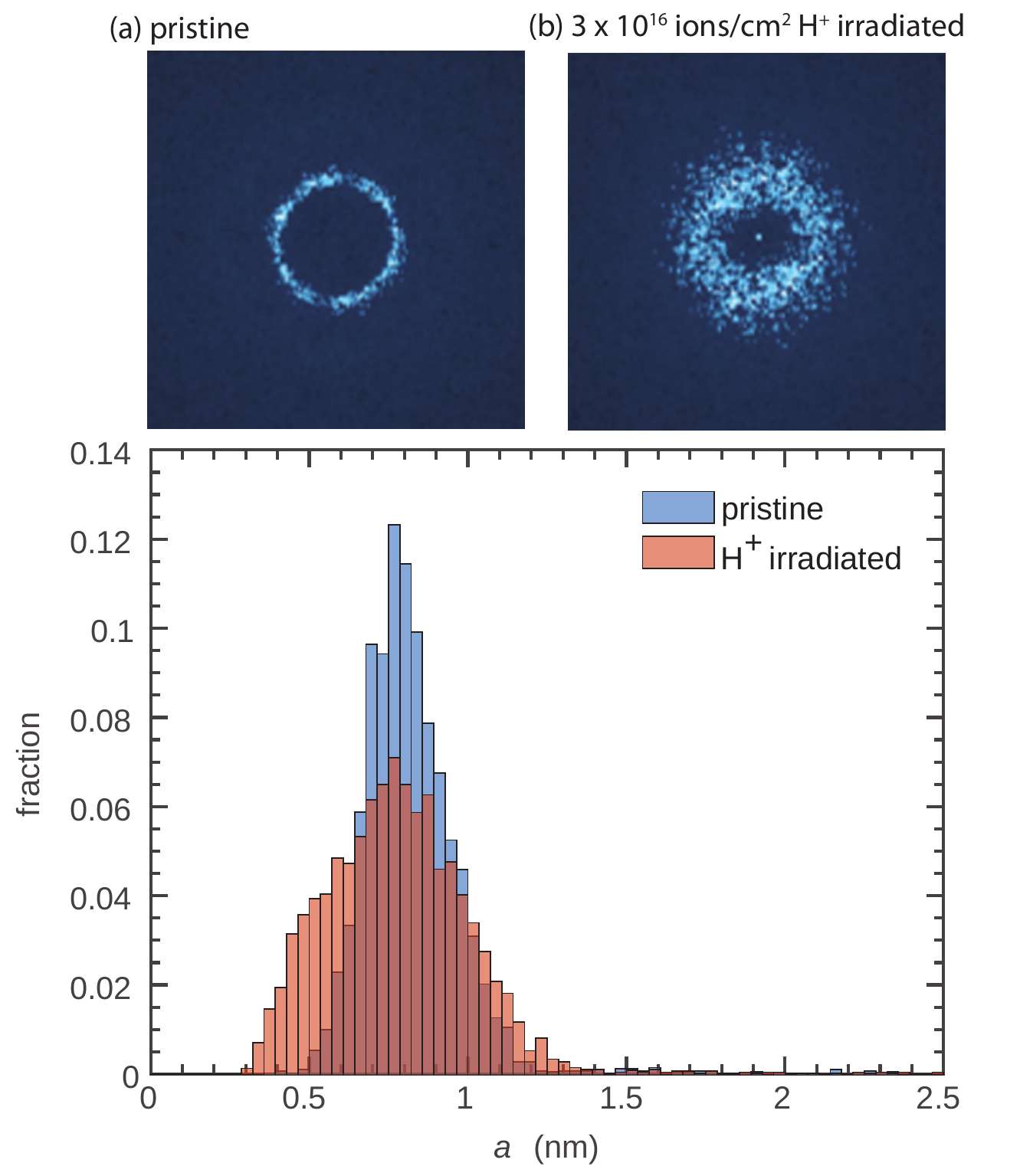}
  \end{center}
\caption{ The 2D fast Fourier transforms of Bitter images of Fig. \ref{bitterimage} where (a) is the pristine and (b) is the $3.0 \times 10^{16}$ ions/cm$^2$ proton-irradiated \ sample. (c) Histogram of normalized distribution of edge lengths of Delaunay triangles ($a$) for both irradiated and pristine samples.
}
\label{FF}
\end{figure}

\begin{figure}[b]
\center
  \begin{center}
\includegraphics[width=9 cm]{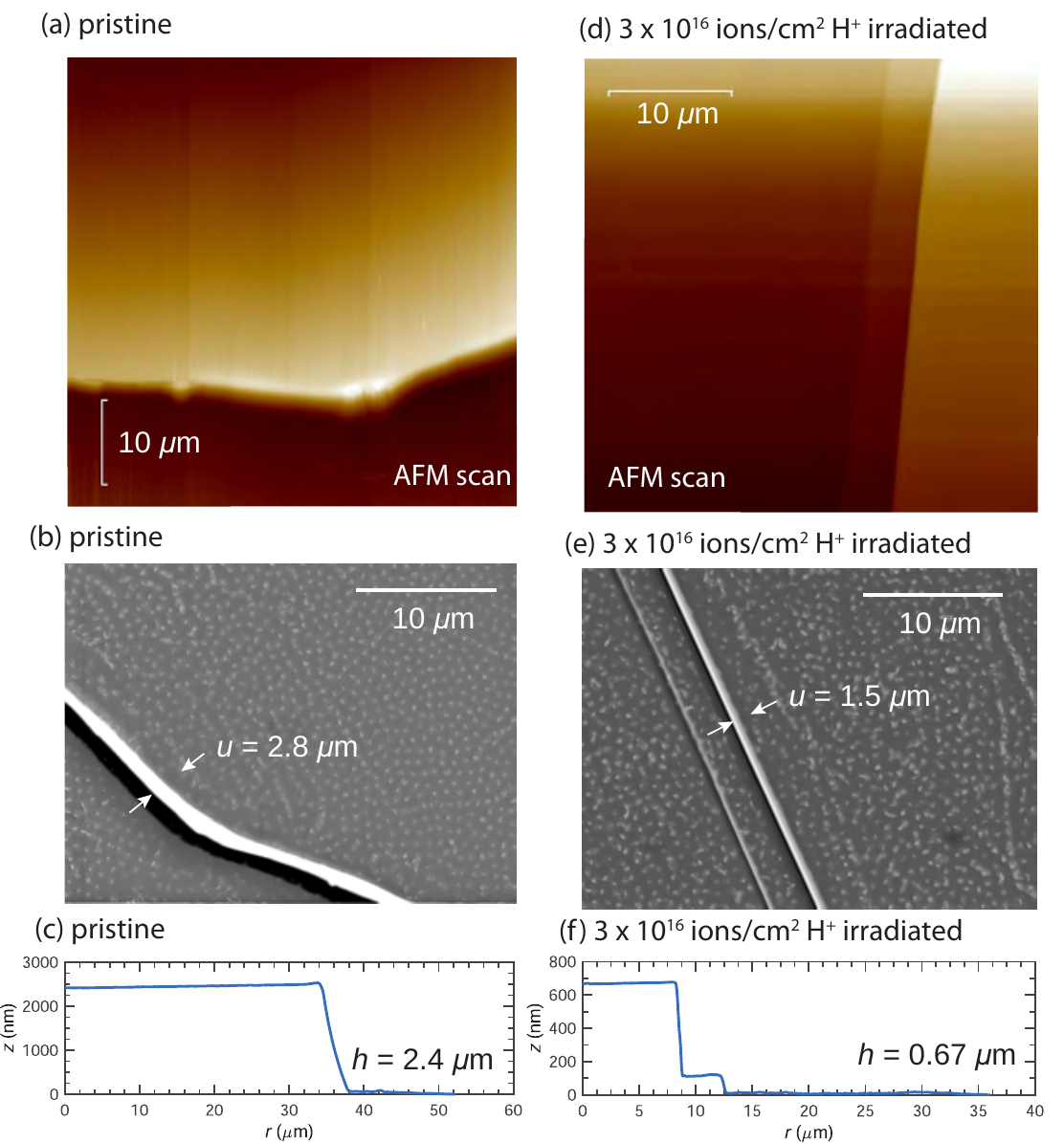}
  \end{center}
\caption{ Steps found on crystal surface of decorated samples to determine the vortex freezing temperature.  (a) and (d) indicate the atomic force microscope (AFM) scan of the step of the pristine and $1 \times 10^6$ ions$/$cm$^2$ H$^+$-irradiated crystal, respectively.   (b) and (e) are Bitter decoration images in which the arrows measure the width of the Meissner belt.  (c) and (f) are the profile of the AFM scan to quantify the height of the step.}
\label{AFM}
\end{figure}

From a low irradiation dose of $0.5\times 10^{16}$ ions/cm$^2$, the field-dependence of $\Jcm$ deviates from the $B^{-5/8}$ dependence, lifting upwards especially at high fields, signifying a shift in the pinning mechanism.  Up to a dose of $1.0 \times 10^{16}$ ions/cm$^2$, there is apparent coexistence of the single-vortex pinning mechanism at intermediate fields, but is later fully masked by a weaker field dependence of $\approx B^{-1/3}$, similarly observed in the case of proton-irradiated Ba$_{1-x}$K$_{x}$Fe$_{2}$As$_{2}$ and \BaCox crystals \cite{PhysRevB.86.094527, 0953-2048-28-8-085003}. Recent simulations by a large-scale time-dependent Ginzburg-Landau theory have recognized the change in the power law field dependence owing to increase in pinning volume by spherical defects
\cite{0953-2048-31-1-014001}. 

\begin{figure*}[t]
\includegraphics[width=\textwidth]{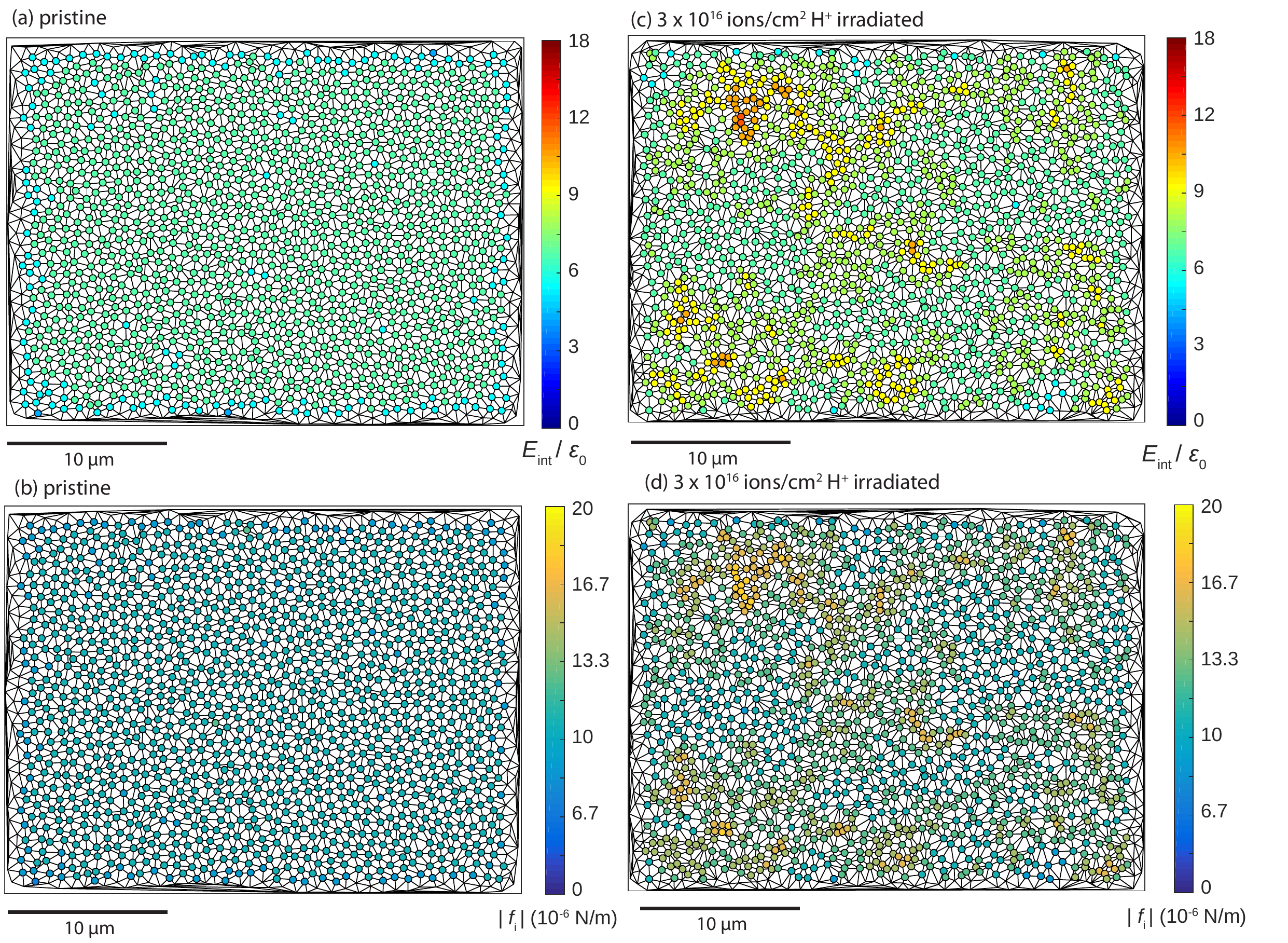}
\caption{Spatial distribution of normalized interaction energy ($E^{int}_i/\varepsilon_0$) for (a) pristine and (c) $3.0 \times 10^{16}$ ions/cm$^2$ H$^+$-irradiated BaFe$_{2}$(As$_{0.67}$P$_{0.33}$)$_2$.  Spatial distribution of pinning force ($|\textbf{f}_i|$) (b) pristine and (d) $3.0 \times 10^{16}$ ions/cm$^2$ H$^+$-irradiated BaFe$_{2}$(As$_{0.67}$P$_{0.33}$)$_2$. }
\label{interaction energy}
\end{figure*}

\begin{figure*}[t]
\includegraphics[width=\textwidth]{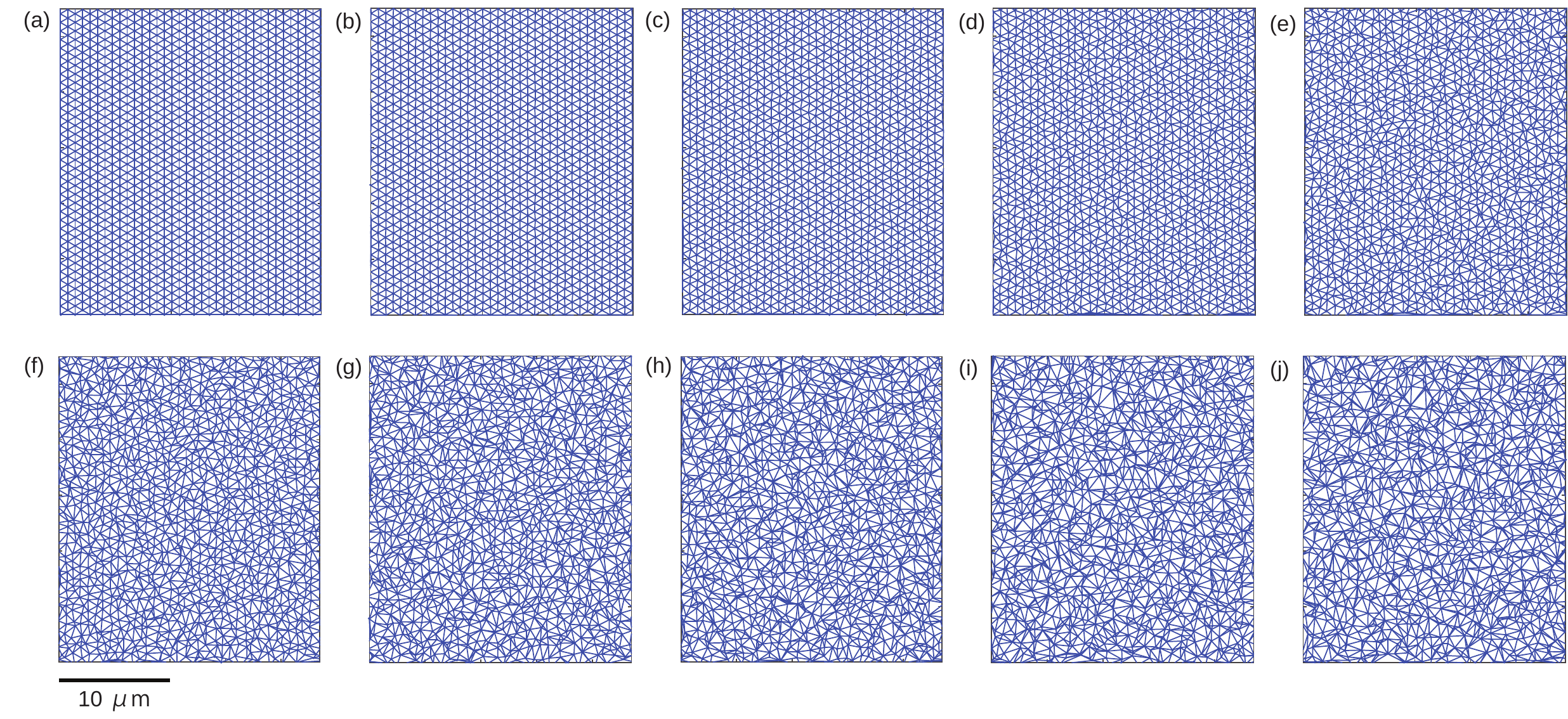}
\caption{Simulated vortex ensemble with varying degree of disorder ($\beta$) : (a) $\beta$ = 0, (b)$\beta$ = 0.05, (c) $\beta$ = 0.1, (d) $\beta$ = 0.15 , (e) $\beta$ = 0.2, (f) $\beta$ = 0.25, (g) $\beta$ = 0.3, (h) $\beta$ = 0.35,  (i) $\beta$ = 0.4 and (j) $\beta$ = 0.45}.
\label{simulation}
\end{figure*}

 Above that, remarkably, the nonmonotonic behavior, referred to as the peak effect present in the pristine state is cloaked through irradiation.  This is natural considering that the peak effect is marked by the crossover from the strong pinning regime to the weak collective pinning regime.  
While the pristine sample marks a crossover, the irradiated sample lacks such a shift owing to the much more firm pinning provided by irradiation prompted defects.
The lack of the fish-tail effect in the sample reported in Ref. \cite{PhysRevLett.105.267002} may signify that the as-grown sample of that investigation has significant defects which prompt strong pinning, thus, masking the weak pinning centers inherent in the crystal.

Yet, upon high irradiation dose, a novel cusplike behavior resembling a distinctive peak effect develops in Figs. \ref{Jc} (d)-\ref{Jc}(f). Note that such a cusplike feature is not unique to this system, as a similar behavior has been reported in Na$_{0.5}$Ca$_{0.5}$Fe$_2$As$_2$ pristine single crystals \cite{PhysRevB.84.094522}.

Looking at the field dependence of the normalized relaxation rate ($d\ln M/d\ln t \equiv S$) as illustrated in Fig. \ref{SH}(a), both in the pristine and proton-irradiated cases, there is a prominent peak in $S$ located at $1 \sim 10$ kOe depending on temperature.  The peak constitutes the increase in vortex motion with increasing field, followed by decrease in vortex motion due to non-negligible vortex-vortex interactions which acts against the vortex motion caused by Lorentz force.  The fact that the location of the peak corresponds to $B^*$ obtained from the field dependence of $\Jcm$ supports the suppression of vortex motion owing to non-negligible intervortex interactions. Concerning the pristine sample, the relaxation rate monotonically decreases after reaching the peak while with the case of proton-irradiated sample, amongst 10 K and 20 K, another rapid increase in the relaxation rate is observed at high fields due to thermally activated plastic motion.  Such rapid flux motion becomes evident at higher temperatures in the proton-irradiated sample due to impairment of superconductivity owing to enhanced quasiparticle excitations with large amount of defects, as conveyed in the decrease in $\Tcm$ (Fig. \eqref{dosedependence}).
As exhibited in Fig. \ref{SH}(b) the local minimum shaped by the peak at $B^*$ and thermally activated giant flux creep corresponds to the location of the cusp-like configuration in $\Jcm$, implying that the two-step behavior in the \Jc emanates by superimposing the positive effect of vortex-vortex interactions and the negative effect of plastic creep.

\subsection{Vortex imaging}

In supplement to bulk magnetization experiments, a surface technique of resolving single vortices through Bitter decoration was employed.  This method directly provides insight on the spatial distribution of vortices, hence, recognizing the combined effect of intervortex interaction and vortex-defect interactions on a microscopic level.  Figure \ref{bitterimage} displays the SEM images of decorated pristine and irradiated sample with the same magnification.  As shown, similar chain-structures of vortices were recognized, as observed in Ref. \cite{PhysRevB.87.094506}.  These vortex chains are asserted to be emanating from the randomly located pinning centers which frustrate the intervortex interactions, thereby elongating the vortex cores. The elongation of the cores as a result breaks symmetry in the system, creating patterns \cite{PhysRevLett.106.137003, PhysRevB.95.104519}. Mapping the single vortices and performing a Delaunay triangulation, vortices with six-fold connectivity are sorted from those with lower or higher degree of connectivity.  The population of vortices which deviate from the six-fold connectivity significantly rises after $3.0 \times 10^{16}$ ions/cm$^2$ H$^+$-irradiation, in which the ratio of number of vortices with six fold to nonsix-fold connectivity increases from $1:0.66$ to $1:1.63$. In percentage, the number of six-fold connectivity decreases from $60 \%$ to $38 \%$ as a consequence of irradiation.

 Accompanying the evolution of the connectivity in the vortex distribution, the two-dimensional fast Fourier transformed (2D FFT) images indicate a circular pattern [Figs. \ref{FF}(a) and \ref{FF}(b)].  The circular pattern is smeared out due to positional and orientational disorder as a consequence of irradiation, asserting that H$^+$ irradiation augments disorder to the vortex ensemble \cite{tinkham1996introduction}.  The average adjacent intervortex distance [i.e. sides of the Delaunay triangles ($a$)] $\braket{a}_\mathrm{pristine} = 0.88$ $\mu$m and $\braket{a}_\mathrm{irradiated} = 0.84$ $\mu$m conform to the lattice constant of an ideal Abrikosov vortex lattice at 40 Oe, $a_0 = (2\Phi_0/\sqrt{3}B)^{1/2} \approx 0.77$ $\mu$m. 

The effective flux density, $B_{\mathrm{eff}}$, was estimated through counting the number of vortices within the SEM frame and using the equation, $ B_{\mathrm{eff}} = N \Phi_0 / A $, in which $N$ is the number of vortices in the frame, $\Phi_0$, is the flux quanta, and $A$ is the area of the frame. For the pristine sample, it was determined that $B_{\mathrm{eff, pristine}} = 35$ G, while for the irradiated sample, $B_{\mathrm{eff, irradiated}} = 40$ G. Upon cooling with a field below $H_{\mathrm{c1}}$ from above $\Tcm$ in the case in which there is no pinning in the sample, the complete expulsion of flux results in a Meissner state in which $B_{\mathrm{eff, pristine}} = 0$ G and a diamagnetic response $4\pi M = -H$ equivalent to the external magnetic field. However, with the presence of pinning, before the vortices are expelled, vortices are locked in place, resulting in a finite $B_{\mathrm{eff}}$, and a reduced magnetization $4\pi M = B_{\mathrm{eff}}-H$. Thus, the higher value of $B_{\mathrm{eff}}$ in the irradiated sample indicates a higher degree of pinning owing to suppressed vortex expulsion.

\vspace{- 10 pt}
\section{Discussions}

Since Bitter decoration was conducted in the field-cooling regime, the vortex position is determined by the force balance between pinning ($f_{\mathrm{pin}}$) and intervortex interaction ($f_{\mathrm{int}}$).  The temperature at which the two forces equate is referred to as the freezing temperature ($T_\mathrm{f}$) or the quenching temperature ie. $f_{\mathrm{pin}}(T_\mathrm{f}) + f_{\mathrm{int}}(T_\mathrm{f}) = 0$.  Since the sample is decorated at liquid He temperature, decoration captures information of the vortex position exactly at the freezing temperature.  Hence, all superconducting parameters such as $\lambda_{ab}$ that correspond to the Bitter images are those at the freezing temperature, thereby making it crucial to determine $T_\mathrm{f}$.

\subsection{Determination of vortex freezing temperature}
Although there are several methods to determine the freezing temperature, such as those suggested by Marchevsky \cite{MARCHEVSKY19972083, doi:10.1134/S0020441219040122}, we employ a method demonstrated by Demirdi\ifmmode \mbox{\c{s}}\else \c{s}\fi{} \textit{et al.}, as it does not require the difficult task of precise experimental temperature control in decoration \cite{PhysRevB.84.094517}.  In this method, we find a step on the surface of a superconductor and observe the vortex distribution on the top surface in the vicinity of the step.  In this region, a single vortex experiences a Bean-Livingston barrier (repulsion by the Meissner current running on the edge and attraction caused by the mirror vortex at the surface) along with line tension, thereby yielding the following relation \cite{PhysRevB.84.094517},
 \begin{align}
B_ae^{-\nu} - B_{int}e^{-2\nu} - \frac{\varepsilon_\lambda^2\varepsilon_0}{\Phi_0}\frac{u\lambda_{ab}}{h}\ln\bigg(\frac{B_{c2}}{2B_{int}}\bigg)& = 0.
\label{BL}
\end{align}
Here, $B_a$ is the applied field which in our case is 40 Oe, $B_{int} = n \Phi_0$ is the magnetic induction, $\nu$ is the parameter representing the vortex-free zone $u$ normalized by $\lambda_{ab}$, and $h$ is the height of the step. $u$ was determined through decoration patterns, and $h$ was quantified through atomic force microscopy (AFM).

As illustrated in Fig. \eqref{AFM}, the step used to determine the freezing temperature is $h_\mathrm{pristine} = 2.4$ $\mu$m, and  $h_\mathrm{irradiated} = 0.67$ $\mu$m for pristine and irradiated samples, respectively.  Moreover, the width of the Meissner belt averaged over several points on the edge,  are $u_\mathrm{pristine} =  2.8$ $\mu$m and $u_\mathrm{irradiated} =  1.5$ $\mu$m.  Using the above experimentally obtained parameters and the force balance equation of Eq. \eqref{BL}, the penetration depth at the freezing temperature was determined to be $\lambda_{ab}(T_\mathrm{f}) \approx 700$ nm for both cases, before and after irradiation.

\subsection{Vortex interaction energy}

The interaction energy of the $i$th vortex with the surrounding vortices ($E^{int}_i$) was determined from the position of vortices using the following equation \cite{PhysRevB.84.094517, PhysRevB.87.094506}
\begin{align}
E^{int}_i & = \sum_j2\varepsilon_0K_0\bigg(\frac{|\textbf{r}_{ij}|}{\lambda_{ab}(T_\mathrm{f})}\bigg).
\label{IE}
\end{align}
The repulsion experienced by the $i$th vortex is given by the sum of the repulsion brought about by all the surrounding vortices which is expressed as a twice the product of the vortex line energy $\varepsilon_0$ and the zero-th order modified Bessel function $K_0$, where $\textbf{r}_{ij}$ is the distance between the $i$th vortex and the surrounding $j$th vortex.
Since vortices at the edge of the frame are less influenced by surrounding vortices, an underestimation of interaction energy occurs. To reduce such edge effect, vortices within 3$\%$ margin from the edge of the frame was omitted from calculations (Fig. \ref{interaction energy}).

\begin{figure}[b]
\center
  \begin{center}
\includegraphics[width=8 cm]{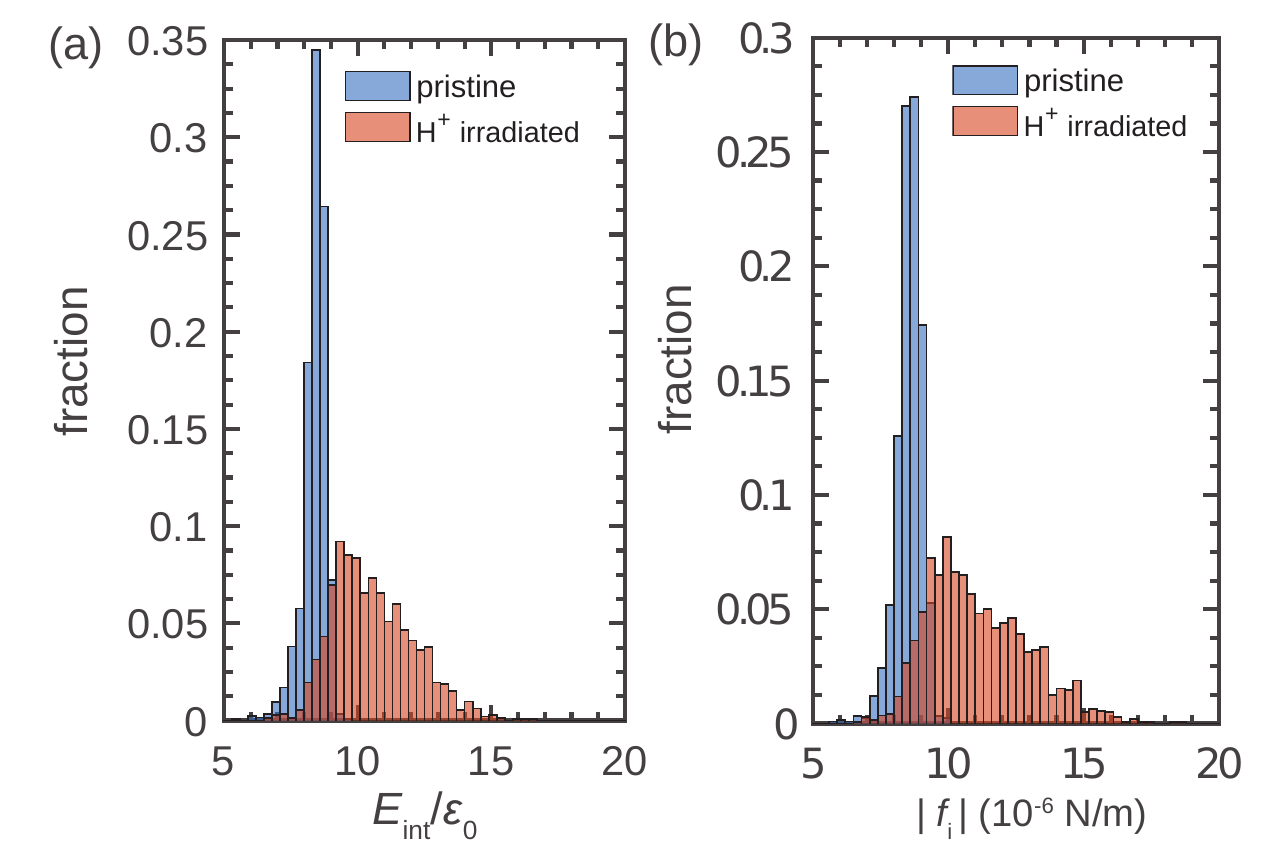}
  \end{center}
\caption{ Histogram of normalized distribution of (a) interaction energy and (b) pinning force of irradiated and pristine BaFe$_{2}$(As$_{0.67}$P$_{0.33}$)$_2$.}
\label{histogram}
\end{figure}

\begin{figure}[t]
\center
  \begin{center}
\includegraphics[width=8.5 cm]{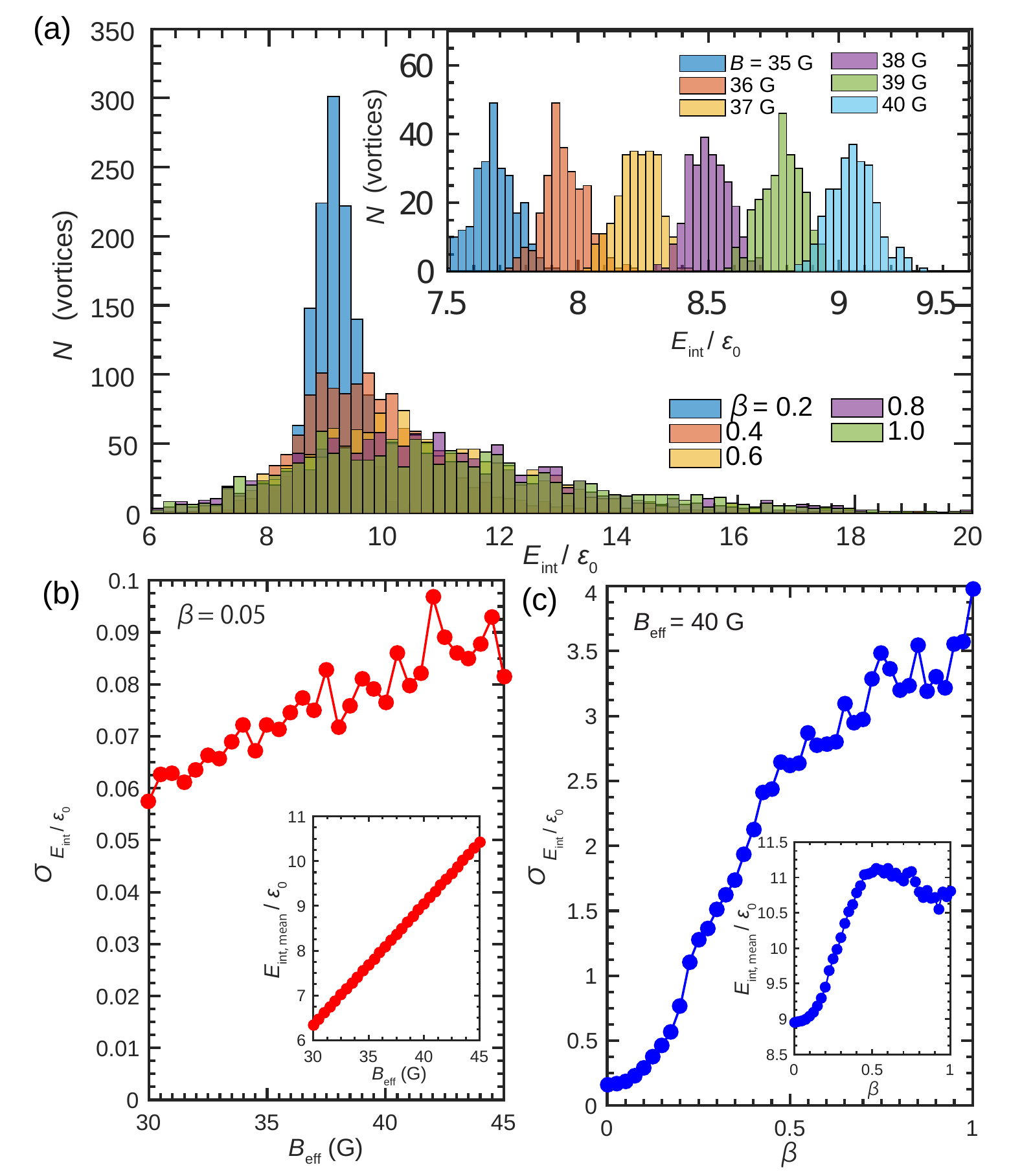}
  \end{center}
\caption{ (a) Histogram of vortex interaction energy of the simulated vortex ensemble upon varying $\beta$ at fixed effective flux density $B_{\mathrm{eff}}=40$ G, and (inset) varying $B_{\mathrm{eff}}$ at fixed  $\beta = 0.05$. (b) The $B_{\mathrm{eff}}$-dependence of the interaction energy standard deviation and (inset) the $B_{\mathrm{eff}}$-dependence of the interaction energy mean. (c) The $\beta$-dependence of the interaction energy standard deviation and (inset) the $\beta$-dependence of the interaction energy mean. }
\label{simulation1}
\end{figure}

The average normalized interaction increases from $E^{int,\mathrm{pristine}}_i/\varepsilon_0 = 8.4 $ to $E^{int,\mathrm{irradiated}}_i/\varepsilon_0 = 10.6$ upon irradiation.  As the Bitter decoration method was performed through field-cooling from room temperature ($\approx 300$ K), the positions of vortices are captured at the freezing temperature, where the vortex-vortex interaction energy and vortex pinning energy equate. Therefore, the interaction energy grants access to the pinning energy signifying that the shift to a larger average interaction energy manifests higher pinning energies present in those that are irradiated by protons.  Note that elevated levels of interaction energy is found where vortices are more densely packed, due to the presence of defects.
In a similar manner, the pinning force per unit length of the $i$-th vortex ($\textbf{f}_i$) is obtained from the following equation, involving the first order modified Bessel function $K_1$ \cite{PhysRevB.84.094517, PhysRevB.87.094506}.
\begin{align}
\textbf{f}_i & = \sum_j\frac{2\varepsilon_0}{\lambda_{ab}}K_1\bigg(\frac{|\textbf{r}_{ij}|}{\lambda_{ab}(T_{f})}\bigg)\frac{\textbf{r}_{ij}}{|\textbf{r}_{ij}|}
\label{FP}
\end{align}
From the spatial distribution of the magnitude of the pinning energy, $|\textbf{f}_i|$, in the pristine case, a rather homogeneous distribution of the pinning energy is found, whereas after irradiation, discernible patches of areas with immense pinning force are visible.  The average pinning force increases from $|\textbf{f}_i^{\mathrm{ pristine}}| = 8.6 \times 10^{-6}$ N$/$m to $|\textbf{f}_i^{\mathrm{ irradiated}}| = 11.1 \times 10^{-6}$ N$/$m, a $1.3$-fold enhancement.

\subsection{Simulation of disorder-controlled vortex ensembles}

To understand the effect of disorder on the pinning force and the intervortex interaction energies of vortex ensembles, simulation with controlled disorder and flux density was subject for investigation. Initially, a hexagonal lattice was formulated to simulate an ideal Abrikosov vortex lattice. From the ideal vortex ensemble, controlled disorder was implemented by perturbing the vortex position by a disorder parameter ranging from zero to unity ($\beta$). The position of each vortex is changed randomly bound by the factor $\beta \times a_0$.  The vortex ensembles with varying degrees of disorder are exhibited in Fig. \ref{simulation}. By fixing the value $\beta = 0.05$, and varying the effective flux density from $B_{\mathrm{eff}}=30$ G to 40 G,  the mean interaction energy increases linearly [Fig. \ref{simulation1}(b) inset]. Yet, the broadening in the interaction energy observed experimentally in Bitter decoration, as exhibited in Fig. \ref{histogram}, cannot be attributed simply by the difference in $B_{\mathrm{eff}}$ before and after irradiation, as the standard deviation in the distribution increases only by 0.04$\varepsilon_0$, while the standard deviation increases by 1.06$\varepsilon_0$, from 0.45$\varepsilon_0$ to 1.51$\varepsilon_0$.

Through increasing $\beta$ and at a fixed value of $B_{\mathrm{eff}} = 40$ G, both the mean interaction energy increases and saturates at at higher degrees of disorder at approximately, $\beta > 0.4$ [Fig. \ref{simulation1}(c)]. In addition, from the histogram of Fig. \ref{simulation1}(a), the distribution flattens out and becomes wider. The standard deviation significantly increases from 0.1 $\varepsilon_0$ to 4.0 $\varepsilon_0$ (Fig. \ref{simulation1}(c)).  Hence, the experimentally observed widening and the increase in the distribution of the pinning energy and the interaction energy is consistent with the fact that the increase in disordered from a vortex ensemble resembles a pristine lattice to a glassier disordered state.

\section{Conclusion} 

An extensive investigation was performed to thoroughly understand the effects of disorder in BaFe$_{2}$(As$_{0.67}$P$_{0.33}$)$_2$. As opposed to previous studies in which the samples studied had apparent spatial distribution in superconductivity, the samples subject here were more spatially homogeneous as indicated through the MO imaging, and less disordered as evident through evaluation from transport measurements.  This allowed for a systematic assessment of the role of irradiation-induced disorder, which was incorporated through 6 MeV proton irradiation. 

Owing to the high quality of pristine \sample crystals synthesized through the Ba$_2$As$_3$/Ba$_2$P$_3$ flux method \cite{doi:10.1143/JPSJ.81.104710}, we are able to observe a fish-tail effect that was not present in previous reports \cite{PhysRevLett.105.267002, PhysRevB.87.094506}. Through irradiation, the fish-tail feature was masked by vortex pinning in irradiation prompted defects. Increasing the dose of irradiated ions, the critical current density increases steadily up to a dose of 10$^{16}$ ions/cm$^2$ and saturates at higher doses. To gain insight into the evolution of the vortex phase, the $\Jcm$ was analyzed through the strong pinning model. While the pristine sample fits well into the strong pinning  picture with $\Jcm$ being proportional to $B^{-5/8}$, the field dependence of $\Jcm$ at higher doses turns into $B^{-1/3}$ dependence \cite{PhysRevB.86.094527, Taen2011784, 0953-2048-28-8-085003}. Moreover, upon irradiation, a unique cusplike feature appears. We discuss that the cusplike feature with positive concavity corresponds to the local minimum of the relaxation rate, representing a vortex phase crossing over from a state of vortex locked by substantial intervortex interactions to a state where flux creep becomes significant.

To further investigate the change in the vortex ensemble prompted by proton irradiation, single vortex imaging by Bitter decoration was performed, revealing that irradiation destroys the ordered triangular lattice into a glassy state. Computing the vortex-vortex interaction energy and the pinning force it was clarified that proton irradiation not only causes shifts to higher energy and force, it causes a  broadening in the overall distribution. These effects resemble that of Co-doping, as opposed to P doping, where such shifts are absent. 

Transport measurements in Co, P, and K doped BaFe$_2$As$_2$ have exhibited that the farther the dopants are from the Fe plane, the smaller the degree of impurity scattering, suggesting that the superconducting currents mainly reside in the Fe plane \cite{doi:10.1021/ja311174e}. Hence, with vortices centered around the Fe plane, pinning by dopants in the Fe plane would have a higher contribution, entailing the shift in vortex interaction energy with increasing Co-doping in the Fe site, as opposed to the absence of shift with increasing P doping in the As site. Since proton irradiation introduces point-defects in random sites, the defect it causes in the Fe site is what could give a pinning behavior resembling Co doping. We suggest that direct evidence of this explanation could be obtained through observing local density of states using a scanning tunneling microscopy around impurities and defects.

\section{Acknowledgements}
This work is partly supported by a Grant-in-Aid for Scientific Research (A) (17H01141)  from JSPS.


%

\end{document}